\documentclass[preprint2]{aastex}
\usepackage{natbib}

\def\aap{Astron. Astrophys.}
\def\apj{Astrophys.~J.}
\def\jrasc{J.~R.~Astron. Soc. Can.}

\newcommand{\ms}{M$_{\odot}$}
\newcommand{\rs}{R$_{\odot}$}

\newcommand{\ha}{H$\alpha$ }

\shorttitle{Modeling of temporal $V/R$ variations in binaries}
\shortauthors{Chadima et al.}

\begin{document}

\title{Hydrodynamical and radiative modeling of temporal \ha emission $V/R$ variations caused by a discontinuous mass transfer in binaries}

\author{Pavel~Chadima\altaffilmark{1}, Roman~Fi\v{r}t\altaffilmark{2}, Petr~Harmanec\altaffilmark{1}, Marek~Wolf\altaffilmark{1}, Domagoj~Ru\v{z}djak\altaffilmark{3},
Hrvoje~Bo\v{z}i\'c\altaffilmark{3}, Pavel~Koubsk\'y\altaffilmark{4} \email{pavel.chadima@gmail.com}}

\altaffiltext{1}{Astronomical Institute of the Charles University, Faculty of Mathematics and Physics, V Hole\v sovi\v ck\'ach 2, CZ-180 00 Praha 8, Czech Republic}
\altaffiltext{2}{Mathematical Institute, University of Bayreuth, D-95447 Bayreuth, Germany}
\altaffiltext{3}{Hvar Observatory, Faculty of Geodesy, University of Zagreb, 10000 Zagreb, Croatia}
\altaffiltext{4}{Astronomical Institute of the Academy of Sciences, CZ-251~65~Ond\v{r}ejov, Czech Republic}

\begin{abstract}
\ha emission $V/R$ variations caused by a discontinous mass transfer in interacting binaries with a rapidly rotating accreting star are modelled qualititatively for
the first time. The program ZEUS-MP was used for a non-linear 3-D hydrodynamical modeling of a development of a blob of gaseous material injected into an orbit around a star.
It resulted in the formation of an elongated disk with a slow prograde revolution. The LTE radiative transfer program SHELLSPEC was used to calculate the \ha profiles
originating in the disk for several phases of its revolution. The profiles have the form of a double emission and exhibit $V/R$ and radial velocity variations. However,
these variations should be a temporal phenomenon since imposing a viscosity in given model would lead to a circularization of the disk and fading-out of given variations.
\end{abstract}

\section{Introduction} \label{intro}
The existing modeling of the $V/R$ variations was based almost exclusively on the model of a slowly revolving elongated envelope around a single star, identified in physical
terms with one-armed oscillations \citep[see e.g.][]{Struve1931, Johnson1958, McLaughlin1961, Okazaki1997, Firt2006} in the disk. The envelope was assumed to originate from an
equatoreal mass outflow due to a critical rotation of the star. In this paper, we present the first preliminary investigation of an alternative idea that an elongated envelope
around a star originates from a discontinuous and short-time mass transfer from a companion in a binary system. Given transfer may occur in eccentric binaries during a periastron
passage. Moreover, such an inflow of a material could also be caused by a density enhancement in a stellar wind from a secondary in a form of coronal mass ejections from
a solar-like, chromospherically active secondary.

In Sect.~\ref{hydrodynamic}, we present a hydrodynamical modeling of a discontinuous mass transfer. Radiative modeling of an \ha profile orignating in a formed disk around
an accreting star and a measurement of its $V/R$ and radial velocity (RV hereafter) variations is presented in Sect.~\ref{radiation}.

\section{Hydrodynamical modeling of a discontinuous mass transfer in a binary} \label{hydrodynamic}
We decided to carry out the first modeling of $V/R$ changes imposed by a discontinous mass transfer presented as a blob of a gaseous material put into an orbit around a rapidly
rotating star having a quadrupole term in its gravitational potential. As detailed below, we are using several simplifications which can be challenged but we do not think that
they seriously affect the results on the qualitative basis. Probably the main simplification is the assumption of an inviscid gas. A viscosity of an orbiting gas, which is not
included in this first model, is expected to destroy any asymetry in the disk and all observed effects should be only temporal. Another simplification which could be criticized
is the assumption of an optically thin environment. To some defense, we would like to mention that in the early stages of the attempts to model the $V/R$ variations,
\citet{Huang1973} used a model based on the assumption of the optically thin envelope while \citet{Kriz1976} did a similar study assuming optically thick envelopes. While
the K\v{r}\'i\v{z}'s line profiles look more realistic, both authors obtained a reasonable description of the $V/R$ changes since the velocity field and the asymmetry of
the envelope were decisive ones. On the other hand, we should emphasize that we make {\sl no assumptions} about the disk in our attempt and gradually build it via a non-linear
hydronamical modeling of an evolution of a discountinuous mass inflow.

\begin{figure*}
\centering
\includegraphics[angle=270,width=0.34\hsize]{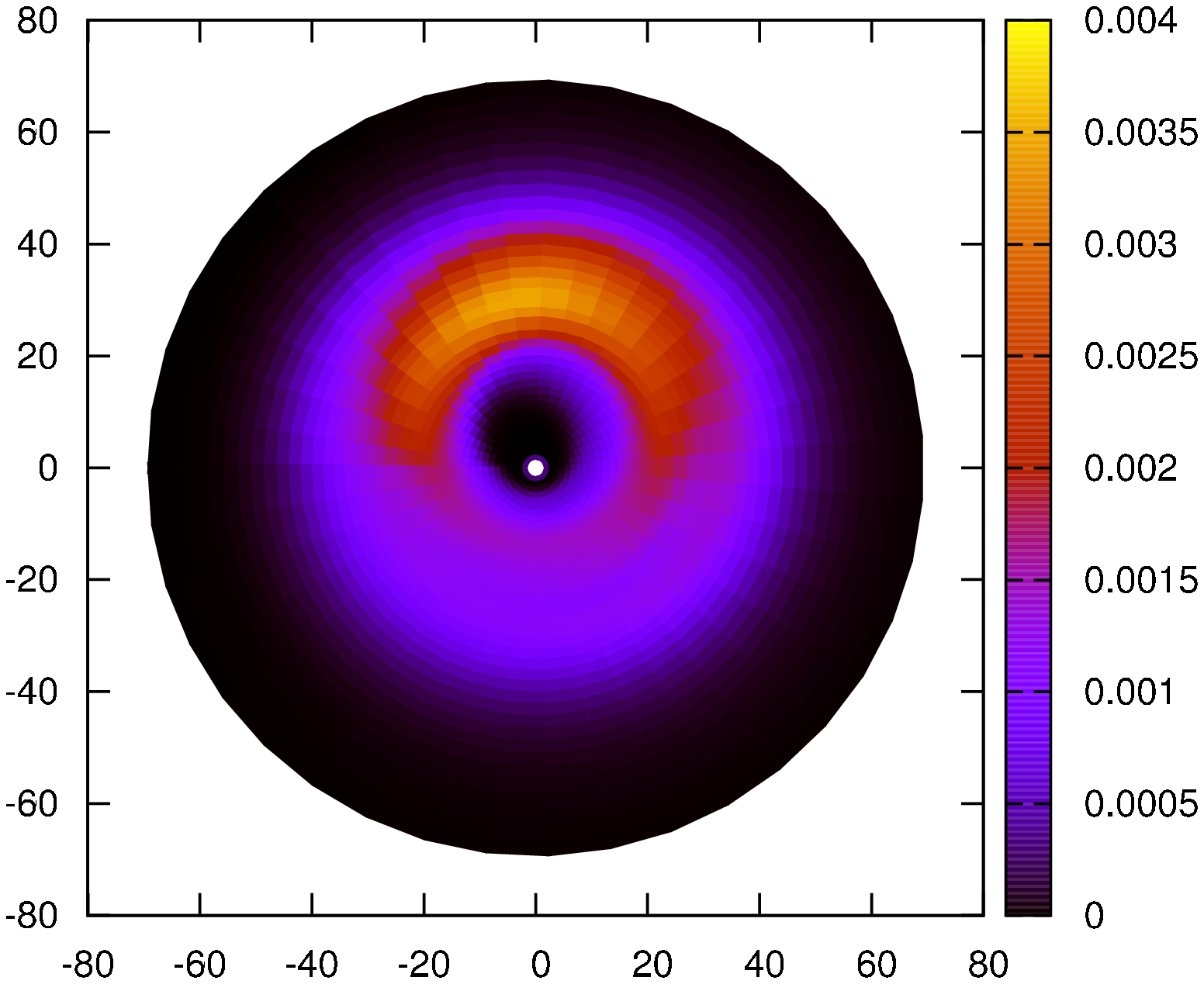}
\includegraphics[angle=270,width=0.27\hsize]{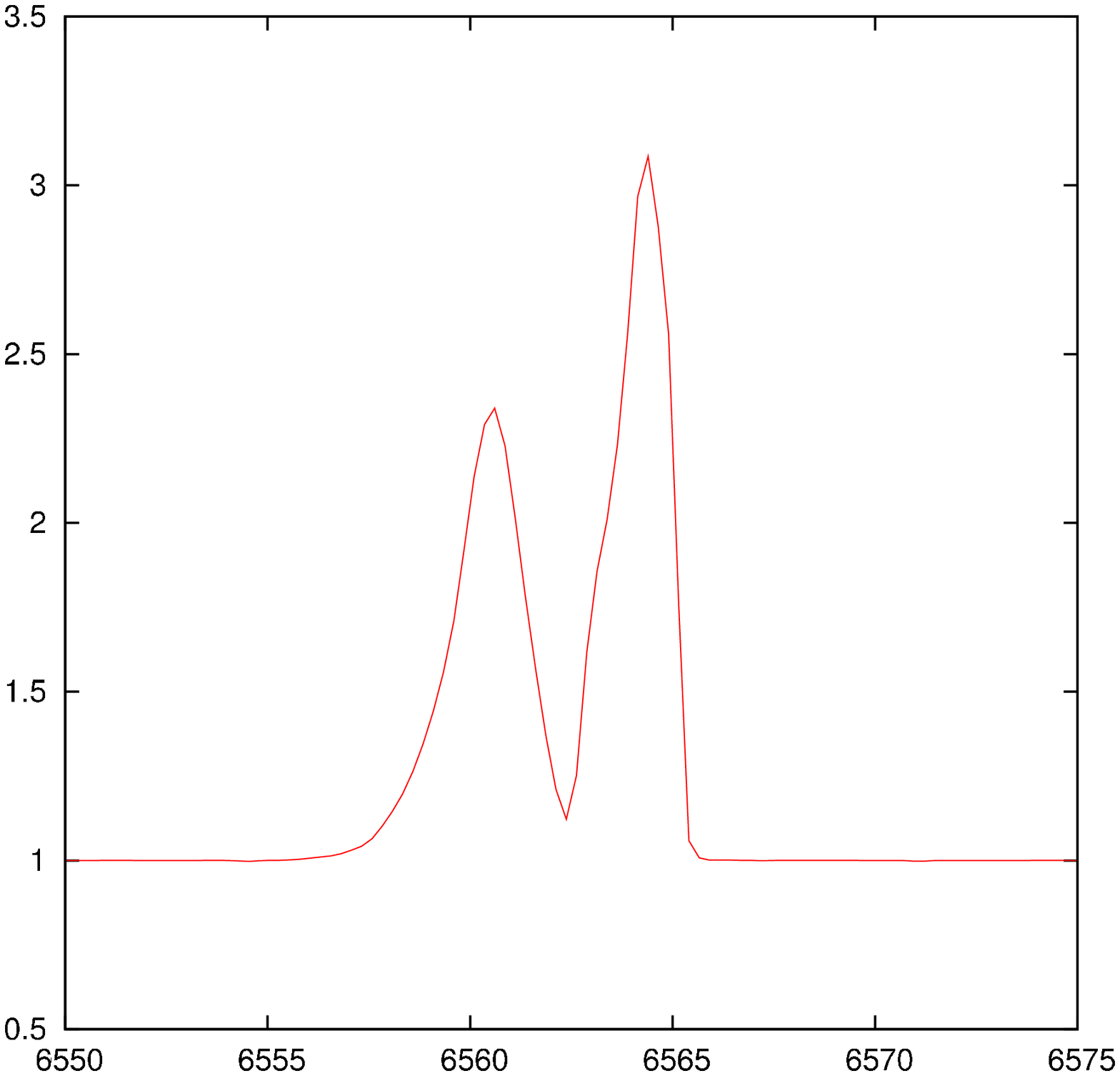}\\
\includegraphics[angle=270,width=0.34\hsize]{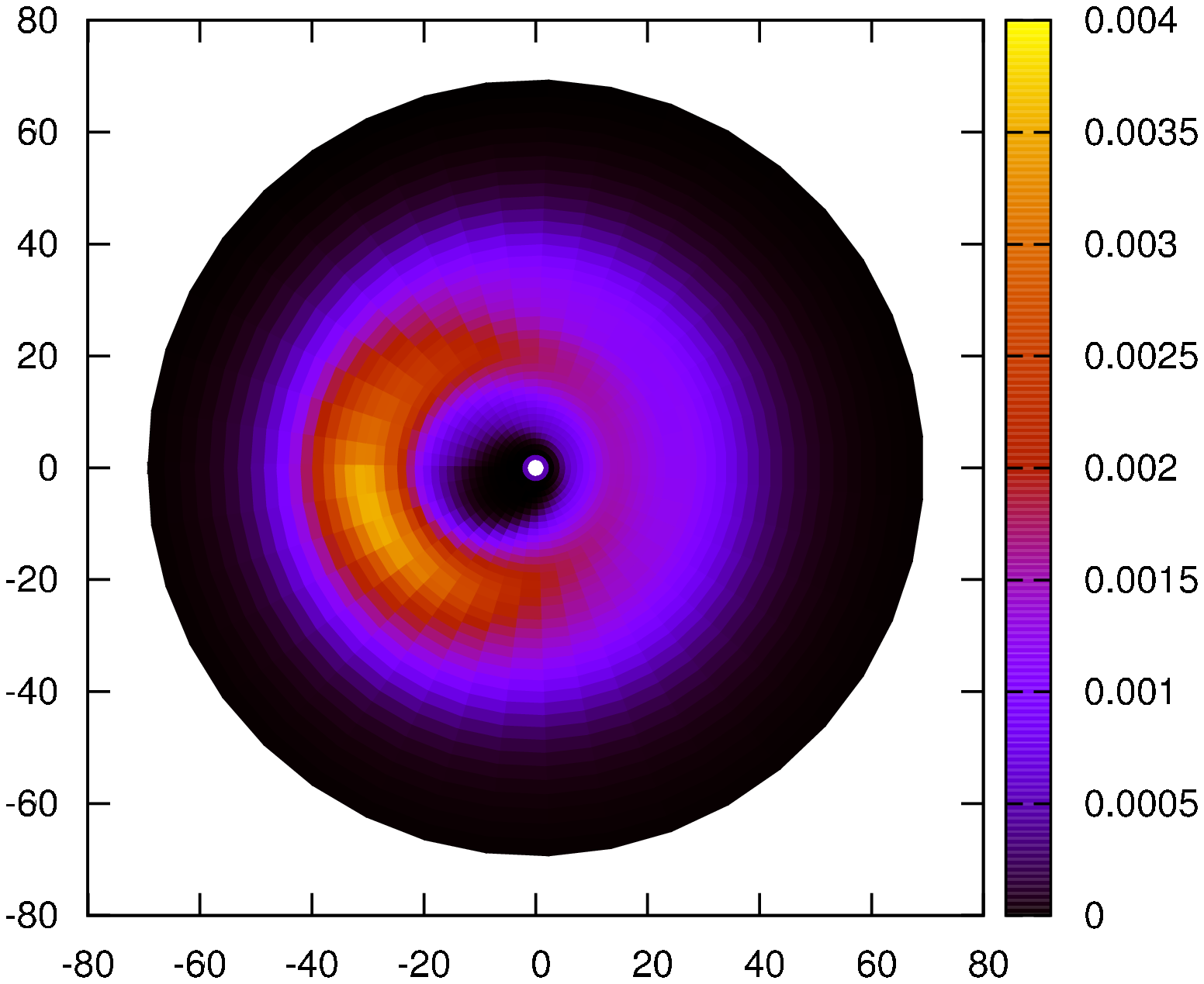}
\includegraphics[angle=270,width=0.27\hsize]{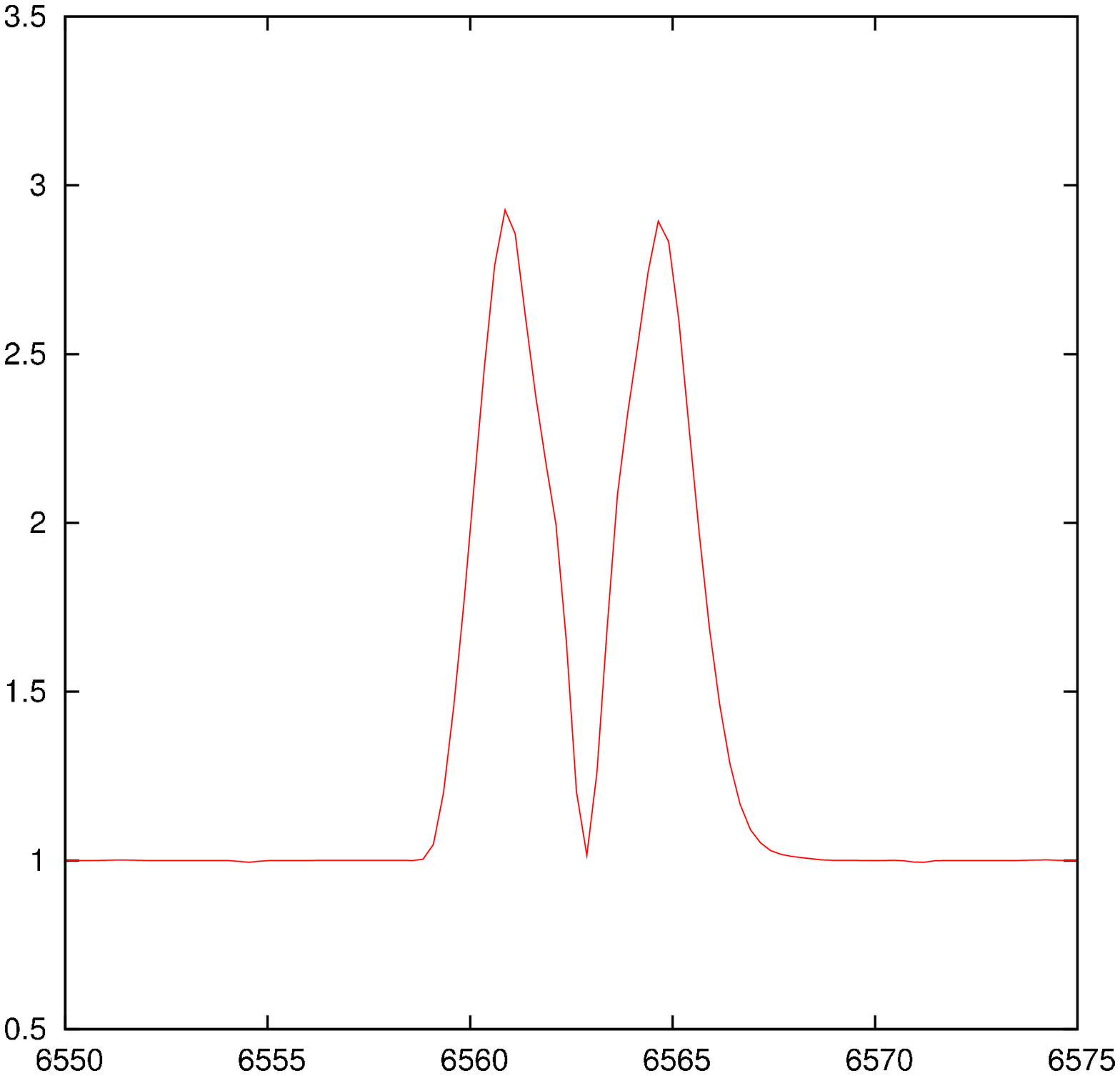}\\
\includegraphics[angle=270,width=0.34\hsize]{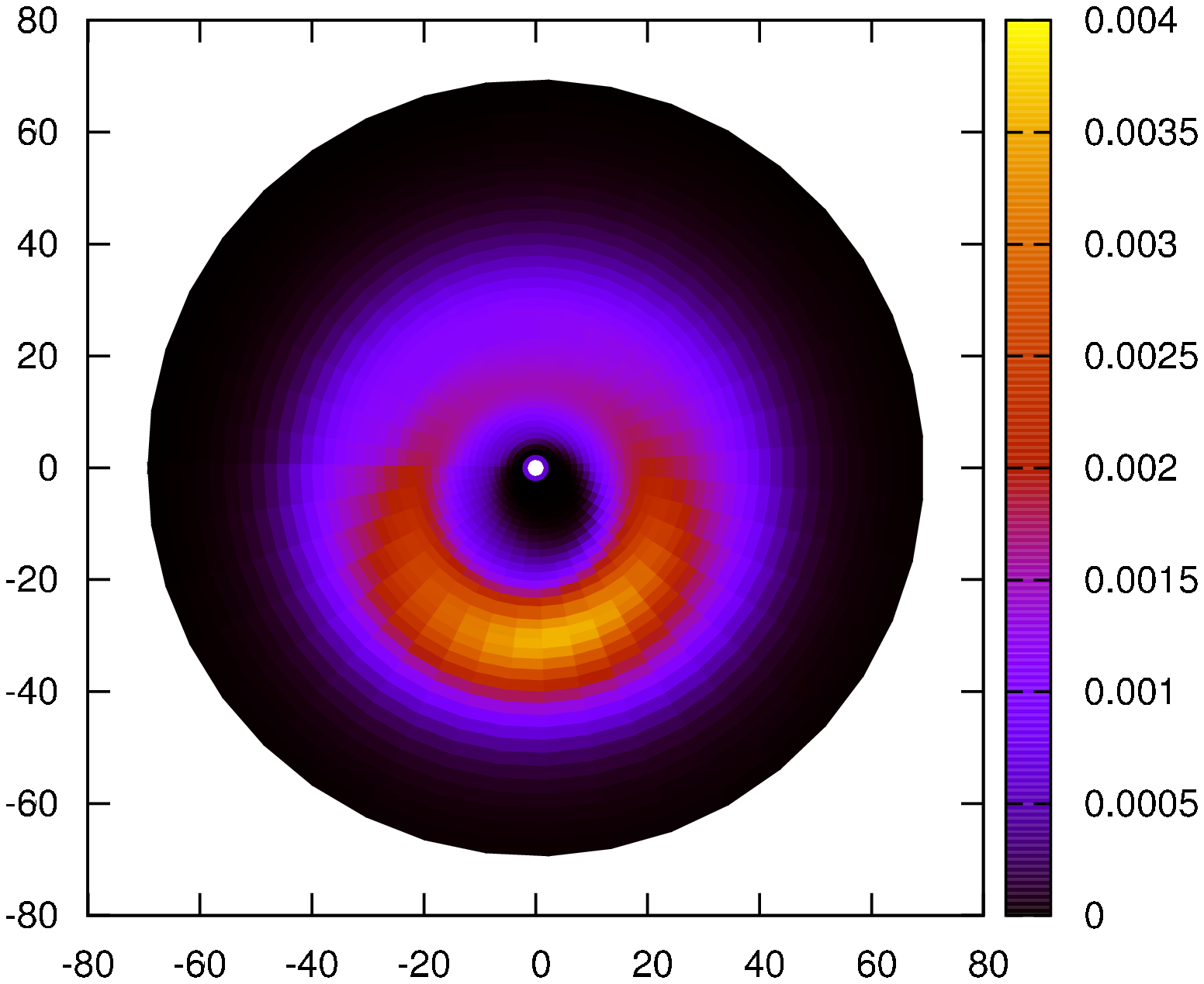}
\includegraphics[angle=270,width=0.27\hsize]{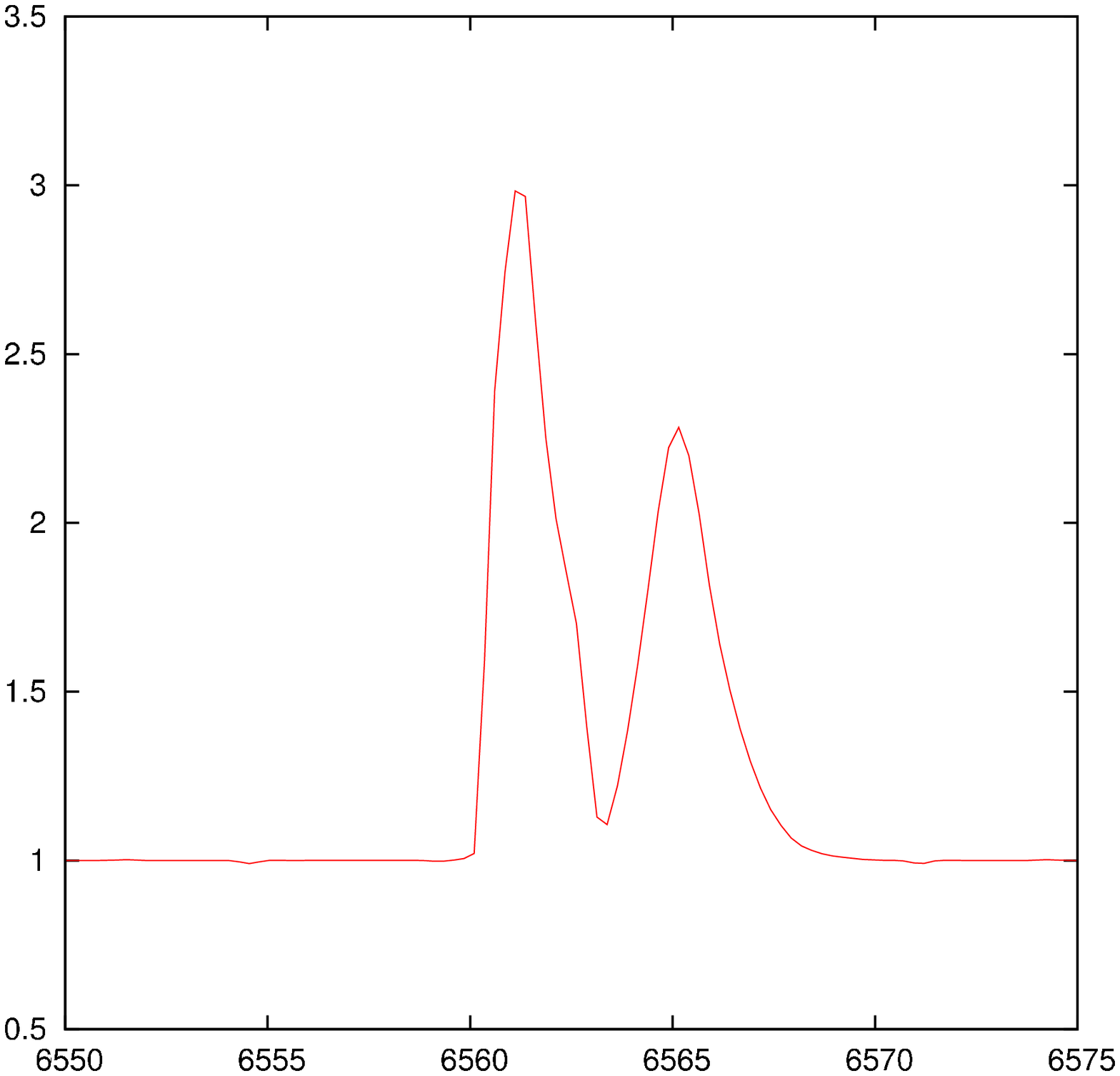}\\
\includegraphics[angle=270,width=0.34\hsize]{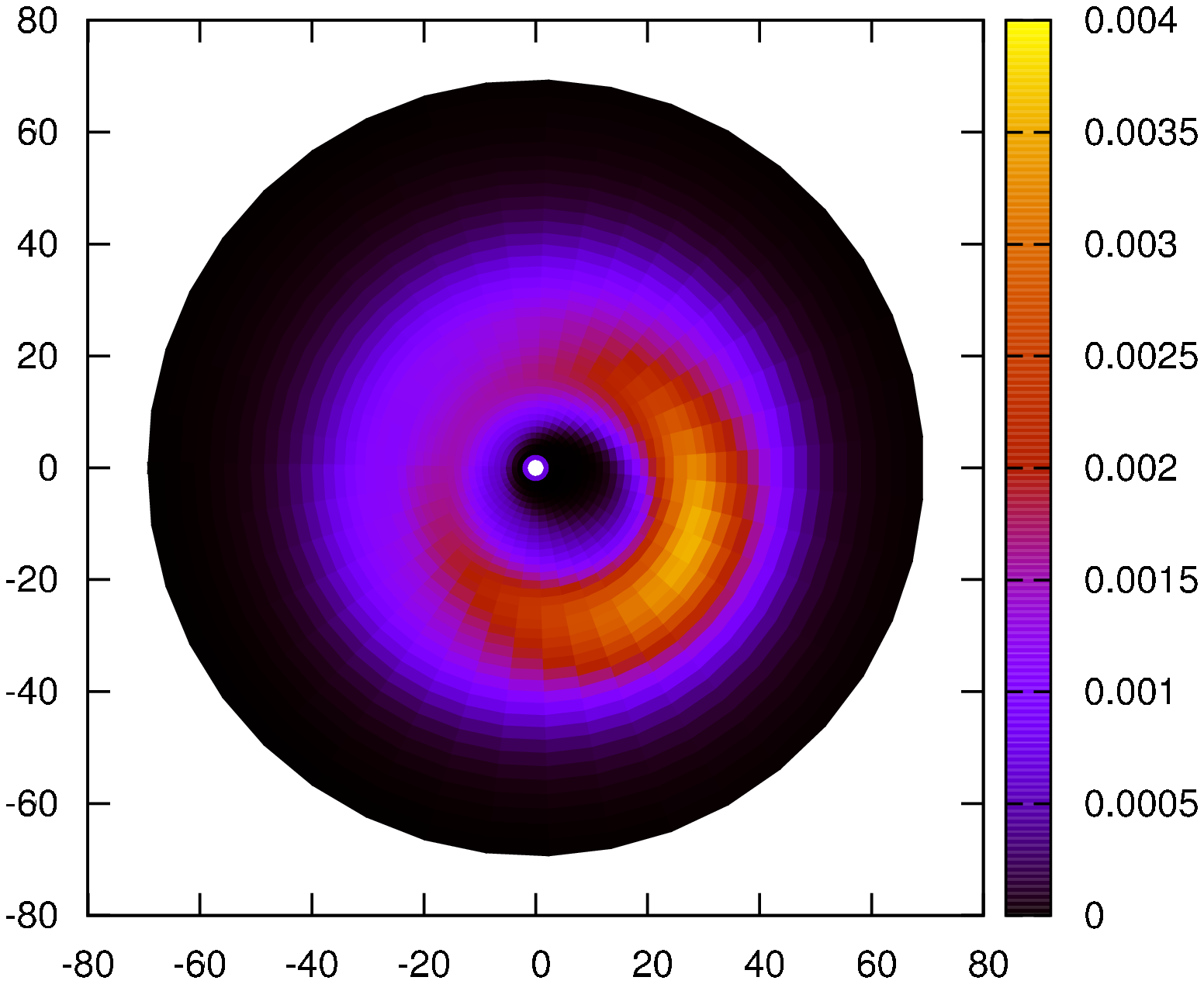}
\includegraphics[angle=270,width=0.27\hsize]{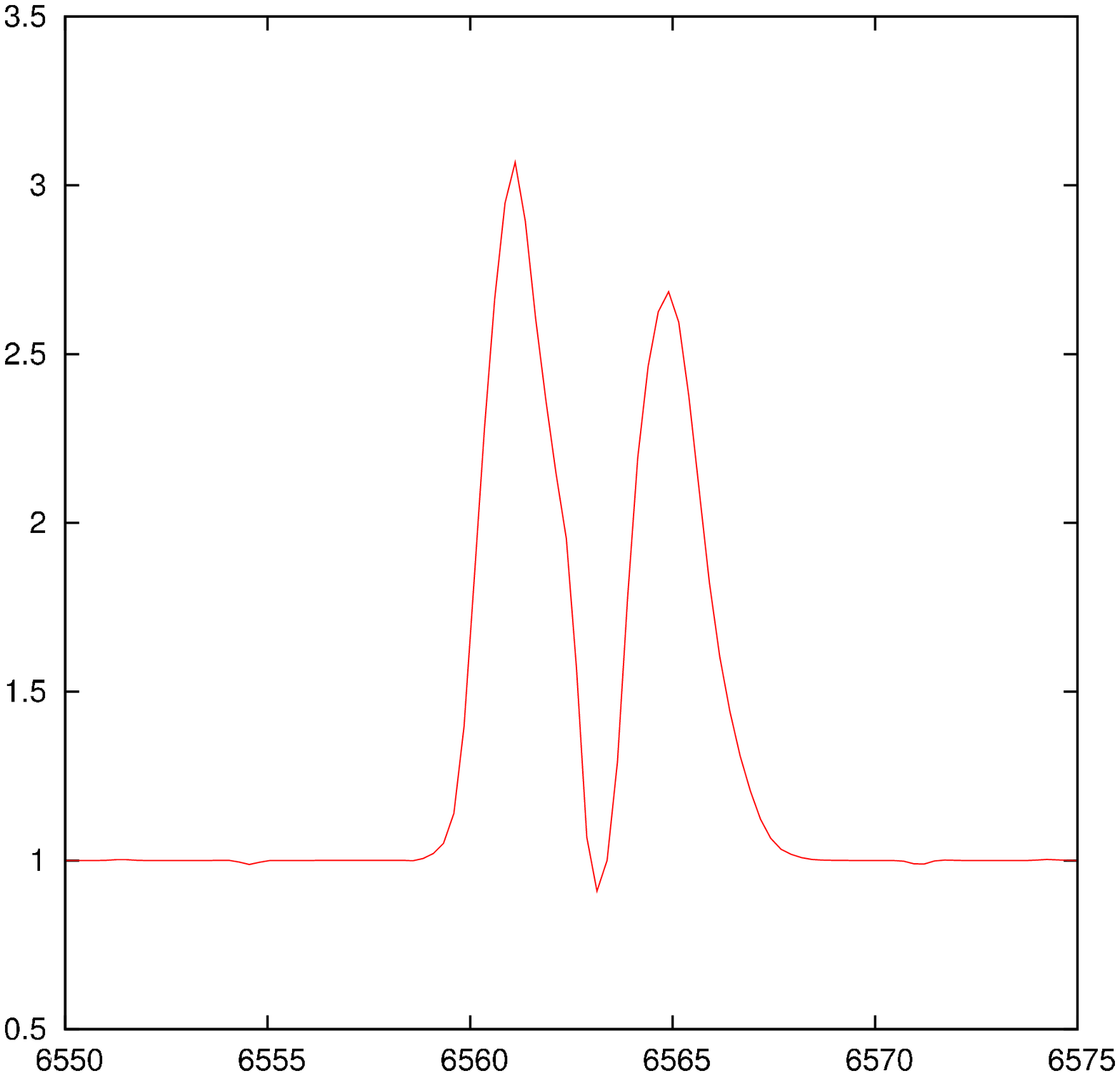}
\caption{A density structure in an equatorial plane of the disk (left) and the \ha line profiles resulting from it (right) for various phases of the prograde rotation of the
elongated density enhancement. Gas particles in the disk are orbiting counter-clockwise and a direction towards an observer is to the right along a horizontal line. See the
electronic edition of the Journal for a color version of this figure. The left side figures are also available as an mpg animation in the electronic edition.}
\label{model}
\end{figure*}

To simulate a hydrodynamical development of a blob of mass originated from a discontinous mass transfer from a secondary, we used the program ZEUS-MP, a multiprocessor clone of
the original program ZEUS-3D \citep{Vernaleo2006}. A computational domain was a hollow cylinder which had the following dimensions in cylindrical coordinates $(z,r,\varphi)$:
\begin{eqnarray}
-0.1R_{\ast} & \le z \le & 0.1R_{\ast}, \nonumber \\
R_{\ast} & \le r \le & 70R_{\ast}, \nonumber \\
0 & \le \varphi \le & 2\pi. \nonumber
\end{eqnarray}
To its centre, we placed a rapidly rotating star with the mass $M_{\ast} = 11$~\ms\ and the radius $R_{\ast} = 5.5$~\rs\ with a gravitational potential $\Phi(z,r,\varphi)$ given by
\begin{eqnarray}
\Phi = \frac{-GM_{\ast}}{\sqrt{r^{2}+z^{2}}} \left[1 + \frac{k_{2}f^{2}}{3} \frac{R_{\ast}^{2}}{r^{2}+z^{2}} \left(1 - \frac{3z^{2}}{r^{2}+z^{2}} \right) \right], \nonumber
\end{eqnarray}
where $k_{2}$ denotes an apsidal motion constant and $f$ is a ratio of a surface rotation to a critical Keplerian rotation:
\begin{eqnarray}
f = \frac{\Omega(R_{\ast})}{\Omega_{\mathrm{K}}(R_{\ast})}. \nonumber
\end{eqnarray}
For our simulation, we chose these parameters as $k_{2} = 0.03$ and $f = 0.95$. A circumstellar matter consists of inviscid, non-selfgravitating ideal gas with {\sl a constant
temperature} $T_{\mathrm{d}} = 10000~K$. The dynamical evolution of the gas is described by standard hydrodynamical equations for an isothermal case
\begin{eqnarray}
\frac{\mathrm{D}\rho}{\mathrm{D}t} + \rho \nabla \cdot \mathbf{v} & = & 0, \nonumber \\
\rho \frac{\mathrm{D}\mathbf{v}}{\mathrm{D}t} & = & -\nabla p - \rho \nabla \Phi, \nonumber
\end{eqnarray}
where $\rho,p, \mathbf{v}$ and $\Phi$ are the gas density, pressure and velocity and the gravitational potential. The $\frac{\mathrm{D}}{\mathrm{D}t}$ denotes the Lagrangian
derivative. Throughout the simulation, we used dimensionless scaled units (denoted with tildes) for length $l$, mass $m$ and time $t$. These are related to physical units as
follows:
\begin{eqnarray}
\tilde l := \frac{l}{R_{\ast}}, \tilde m := \frac{m}{M_{\ast}}, \tilde t := \sqrt{\frac{GM_{\ast}}{R^{3}_{\ast}}} t, \nonumber
\end{eqnarray}
where $G$ denotes the universal gravitational constant. Note that with this choice we have $\tilde G=1$ and the computational domain has the dimensions
$-0.1 \le \tilde z \le 0.1,\ 1 \le \tilde r \le 70,\ 0 \le \tilde \varphi \le 2\pi$.

As initial conditions for density and velocity components we impose
\begin{eqnarray}
\tilde \rho = 10^{-4}, \tilde v_{z} = \tilde v_{r} = 0, \tilde v_{\varphi} = \tilde r^{-3/2} \nonumber
\end{eqnarray}
where $\tilde v_{\varphi}$ corresponds a keplerian velocity. As boundary conditions, we imposed periodic boundary conditions in $\varphi$ and $z$ direction. For an inner boundary
condition at the surface of the star ($\tilde r=1$) we took a slip boundary condition $v_{r} = v_{z} = 0, v_{\varphi} = f$ with $f$ defined above. Regarding an outer boundary
condition, we used a pressure free outflow boundary condition.

To simulate the idea of a short discontinuous mass inflow to an accreting star, we injected a compact ($\tilde r_{\mathrm{inj}} \in [40,42]$, $\tilde \varphi_{\mathrm{inj}} \in
[0,0.2]$, $\tilde z_{\mathrm{inj}} \in [-0.1,0.1]$) blob of mass ($\tilde \rho = 1$) at 80\% of the Keplerian velocity ($\tilde v_{z} = \tilde v_{r} = 0,\tilde v_{\varphi} = 0.8
\tilde r^{-3/2}$) into an initial "interstellar vacuum" and followed its evolution in time.

Our simulation led to a formation of an elongated accretion disk around the central star. A density structure in an equatorial plane of the disk and its time evolution is shown
on the left side of Fig.\,\ref{model} where are several selected frames from a hydrodynamical animation presented in the electronic edition. Material in the disk is orbiting
counter-clockwise and a direction to an observer in all panels is to the right along a horizontal line. It is seen that a denser region is formed near an apocentrum of the
elliptical disk due to a crowding of the orbiting material there. The elongated disk is undergoing a slow prograde rotation with a period of about 16 years.

The prograde rotation of the disk is caused by the quadrupole term of the gravitational potential. This result is in a qualitative agreement with several papers which studied
an influence of the quadrupole term of the gravitational potencial of fast rotating stars on a precession rate of disks around them \citep[see e.g.][]{Savonije1993, Okazaki1996,
Firt2006}. However, the linear, one armed oscilation, numerical simulations presented in these papers use diferent values of input parameter and therefore it is not possible
to do a quantitative comparison of our precesion period with their results. Moreover, it was shown by \citet{Firt2006} that a precesion period is dependent on many free parameters
(one of them is a term $k_{2}f$ which describes a rate of stellar distortion due to its rotation) and the dependence on some parameters is quite strong \citep[see Figs.~3-9 in][]
{Firt2006}. Therefore, by a diferent choise of values of given input parameters, it is possible to obtain a precesion period in a range from $\sim$1~year to several tens of years.

\section{Radiative modeling of the \ha profile} \label{radiation}
To model the emission profiles of the \ha line originating in the disk, we used the program SHELLSPEC \citep{Budaj2004}. It solves a radiative transfer along the line of sight in
an optically thin environment assuming LTE. Only a scattered light from the central star is taken into account, however. The star itself was treated as a black body with a
temperature $T_{\mathrm{s}} = 22000~K$. We chose a view of an observer located in the equatorial plane, i.e. an inclination 90$^\circ$. Resulting profiles of the \ha line for
various phases of the prograde rotation of the disk are displayed on the right side of Fig.\,\ref{model}. The modeling leads to double-peaked emission-line profiles. The denser
region in the disk emits more radiation and its revolution around the central star combined with the rotation of the disk particles results in the $V/R$ changes.

\begin{figure}[t]
\resizebox{\hsize}{!}{\includegraphics{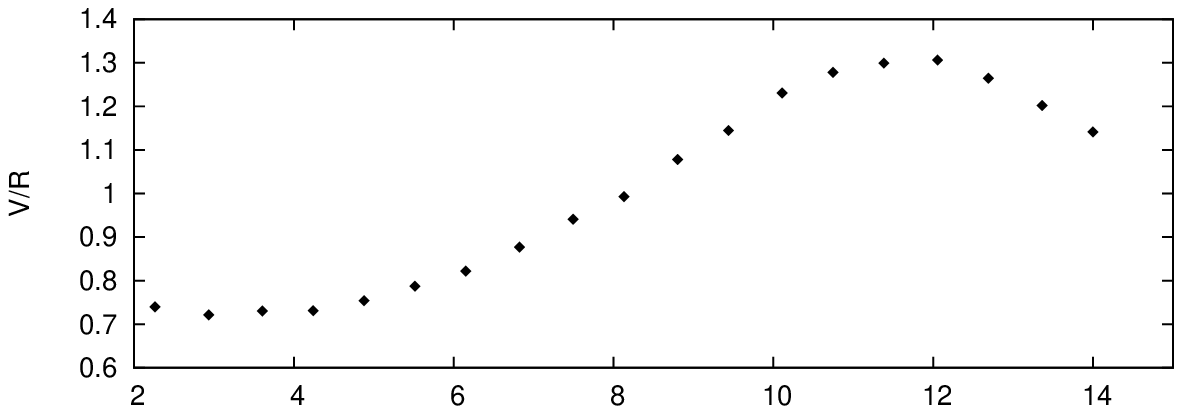}}
\resizebox{\hsize}{!}{\includegraphics{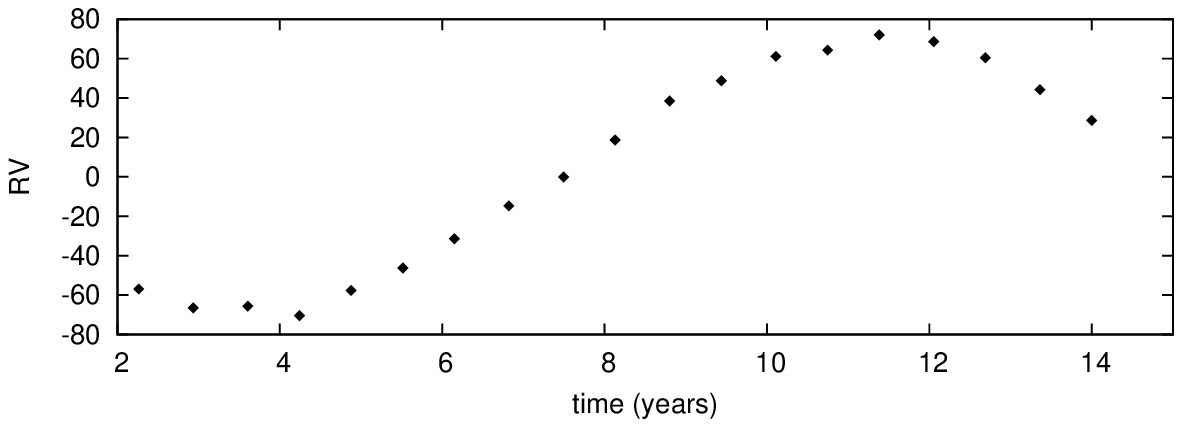}}
\caption{A time variation of the $V/R$ ratios and the RVs for one cycle of the prograde revolution of the disk.}
\label{result}
\end{figure}

We imported a representative selection of the model \ha line profiles into SPEFO \citep{Horn1996, Skoda1996} and measured their $V/R$ ratios and the RVs on wings of the emission.
A time variation of the $V/R$ ratios and the RVs for one cycle of the prograde revolution of the disk is displayed in Fig.~\ref{result}. It is seen that both quantities have a
sinusoidal development and that they are in phase.

\section{Conclusion} \label{conclusion}
We tried to test the idea that a temporal \ha $V/R$ variations can be caused by a discontinuous mass transfer from a companion star in a binary. As an initial simple represantation
of given discountinuous mass transfer, we injected a blob of mass into an orbit around a rapidly rotating star and followed its evolution via 3-D hydrodynamical modeling which led
to a formation of an elliptical disk with a slow prograde revolution. We expect that the similar result would be obtained when injecting a blob of "new" mass into an already
developed circular disk.  A LTE radiative modeling of the \ha line led to the double-peaked emission profile which $V/R$ ratios and RVs undergo a cyclic variations in phase.
However, these variations should be only temporal since a viscosity of an orbiting gas should circularize the disk if there is an absence of a mass transfer from a secondary as
was also shown by \citet{Bisikalo2001}. Therefore also the $V/R$ variations should gradually fade-out.

Nevertheless, we are aware that this first attempt to test given idea was considerably simplified. Beside the already mentioned assumption of the inviscid isotermal gas, the
simplification also involves the way how the discontinuous mass transfer from the companion was represented. Moreover, it is expected that an inclusion of the attractive force
of the orbiting companion (also absent in the present model) could speed up the disk revolution, resulting in a shorter cycle length. Therefore, we intend to improve our modeling
taking into account things mentioned above in some future study.

\acknowledgments
We gratefully acknowledge a use of the programs ZEUS and SHELLSPEC for our modeling. All hydrodynamical computations were carried out on the Sn\v{e}hurka cluster at the Ne\v{c}as
centre for a mathematical modeling in Prague. R.~Fi\v{r}t thanks its administrator M.~M\'adl\'ik for his support. The research of the Czech authors was supported by the grants
205/03/0788, 205/06/0304, 205/08/H005, and P209/10/0715 of the Czech Science Foundation and also from the research project AV0Z10030501 of the Academy of Sciences of the Czech
Republic, from the research plan J13/98: 113200004 of Ministry of Education, Youth and Sports {\sl Investigation of the Earth and Universe} and later also from the Research
Program MSM0021620860 {\sl Physical study of objects and processes in the solar system and in astrophysics} of the Ministry of Education of the Czech Republic. We acknowledge
the use of the electronic bibliography maintained by the NASA/ADS system and by the CDS in Strasbourg.


\begin{thebibliography}{36}
\expandafter\ifx\csname natexlab\endcsname\relax\def\natexlab#1{#1}\fi

\bibitem[{{Bisikalo} {et~al.}(2001){Bisikalo}, {Boyarchuk}, {Kil'Pio}, {Kuznetsov}, \& {Chechetkin}}]{Bisikalo2001}
{Bisikalo}, D.~V., {Boyarchuk}, A.~A., {Kil'Pio}, A.~A., {Kuznetsov}, O.~A., \&  {Chechetkin}, V.~M. 2001, Astronomy Reports, 45, 611

\bibitem[{{Budaj} \& {Richards}(2004)}]{Budaj2004}
{Budaj}, J. \& {Richards}, M.~T. 2004, Contributions of the Astronomical
  Observatory Skalnate Pleso, 34, 167

\bibitem[{{Fi{\v r}t} \& {Harmanec}(2006)}]{Firt2006}
{Fi{\v r}t}, R. \& {Harmanec}, P. 2006, \aap, 447, 277

\bibitem[{{Horn} {et~al.}(1996){Horn}, {Kub\'at}, {Harmanec}, {Koubsk\'y},
  {Hadrava}, {\v{S}imon}, {\v{S}tefl}, \& {\v{S}koda}}]{Horn1996}
{Horn}, J., {Kub\'at}, J., {Harmanec}, P., {et~al.} 1996, \aap, 309, 521

\bibitem[{{Huang}(1973)}]{Huang1973}
{Huang}, S.~S. 1973, \apj, 183, 541

\bibitem[{{Johnson}(1958)}]{Johnson1958}
{Johnson}, M. 1958, Mem. Soc. Royale Sci. Li\`ege IV. Ser, 20, 219

\bibitem[{{K\v{r}\'i\v{z}}(1976)}]{Kriz1976}
{K\v{r}\'i\v{z}}, S. 1976, Bulletin of the Astronomical Institutes of
  Czechoslovakia, 27, 321

\bibitem[{{McLaughlin}(1961)}]{McLaughlin1961}
{McLaughlin}, D.~B. 1961, \jrasc, 13\&55, 73

\bibitem[{{Okazaki}(1996)}]{Okazaki1996}
{Okazaki}, A.~T. 1996, PASJ, 48, 305

\bibitem[{{Okazaki}(1997)}]{Okazaki1997}
{Okazaki}, A.~T. 1997, A\&A, 318, 548

\bibitem[{{Savonije} \& {Heemskerk}(1993)}]{Savonije1993}
{Savonije}, G.~J. \& {Heemskerk}, M.~H.~M. 1993, A\&A,, 276, 409

\bibitem[{{\v{S}koda}(1996)}]{Skoda1996}
{\v{S}koda}, P. 1996, in ASP Conf. Ser. 101: Astronomical Data Analysis
  Software and Systems V, 187

\bibitem[{{Struve}(1931)}]{Struve1931}
{Struve}, O. 1931, ApJ, 73, 94

\bibitem[{{Vernaleo} \& {Reynolds}(2006)}]{Vernaleo2006}
{Vernaleo}, J.~C. \& {Reynolds}, C.~S. 2006, \apj, 645, 83

\end{thebibliography}
\end{document}